# How Brains Are Built: Principles of Computational Neuroscience

By Richard Granger, Ph.D.

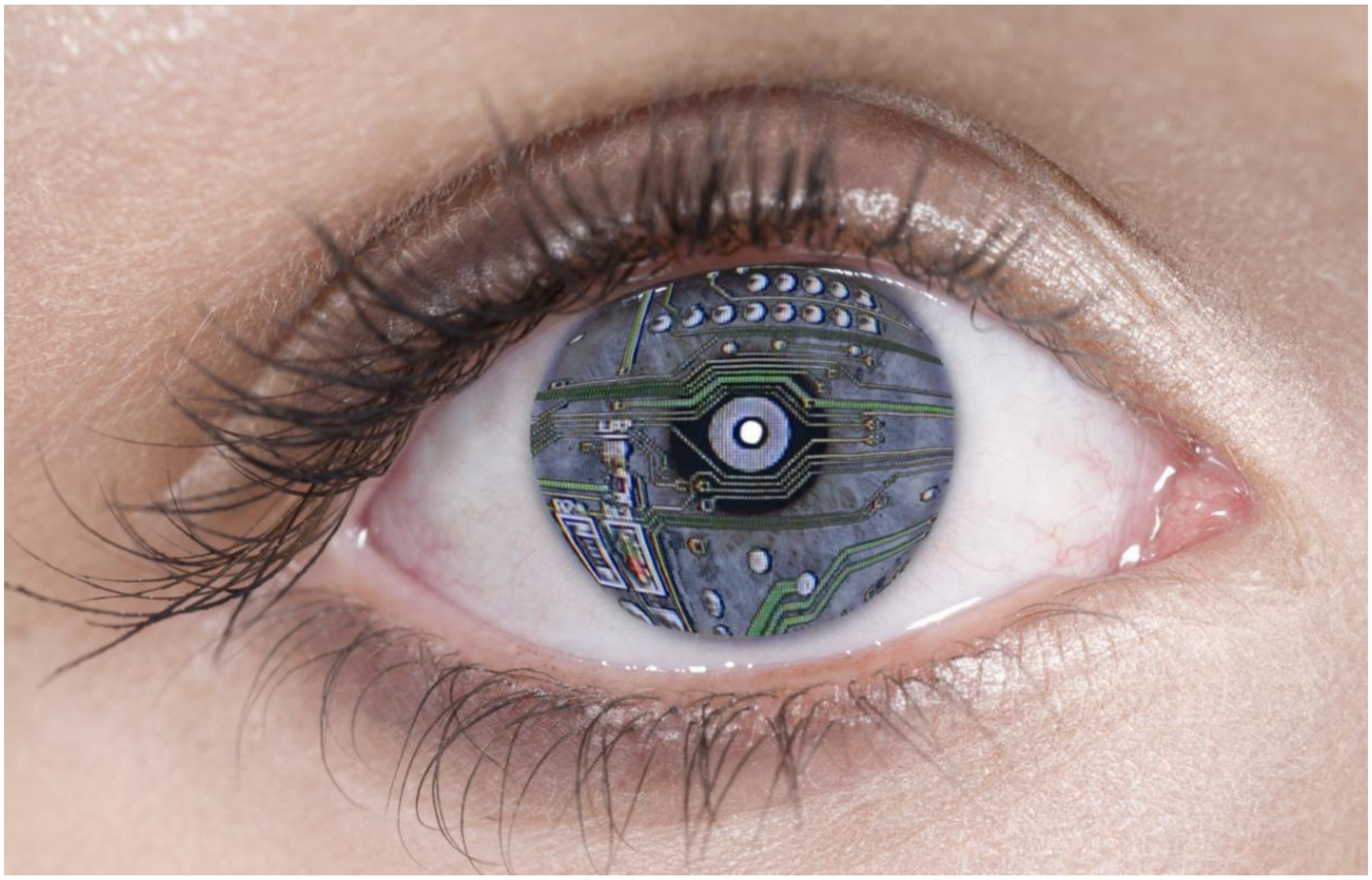

Bernhard Lang/Photographer's Choice/Getty Images

Editor's note: The goal of computational neuroscience is to understand the brain and its mechanisms well enough to artificially simulate their functions. In some areas, like hearing, vision, and prosthetics, there have been great advances in the field. Yet there is still much about the brain that is unknown and therefore cannot be artificially replicated: How does the brain use language, make complex associations, or organize learned experiences? Once the neural pathways responsible for these and many other functions are fully understood and reconstructed, we will have the ability to build systems that can match—and maybe even exceed—the brain's capabilities.

Article available online at http://dana.org/news/cerebrum/detail.aspx?id=30356

"If I cannot build it, I do not understand it." So said Nobel laureate Richard Feynman, and by his metric, we understand a bit about physics, less about chemistry, and almost nothing about biology.[1]

When we fully understand a phenomenon, we can specify its entire sequence of events, causes, and effects so completely that it is possible to fully simulate it, with all its internal mechanisms intact. Achieving that level of understanding is rare. It is commensurate with constructing a full design for a machine that could serve as a stand-in for the thing being studied. To understand a phenomenon sufficiently to fully simulate it is to understand it *computationally*.

"Computation" does not refer to computers per se; rather it refers to the underlying principles and methods that make them work. As Turing Award recipient Edsger Dijkstra said, computational science "is no more about computers than astronomy is about telescopes."[2] Computational science is the study of the hidden rules underlying complex phenomena from physics to psychology.

Computational neuroscience, then, has the aim of understanding brains sufficiently well to be able to simulate their functions, thereby subsuming the twin goals of science and engineering: deeply understanding the inner workings of our brains, and being able to construct simulacra of them. As simple robots today substitute for human physical abilities, in settings from factories to hospitals, so brain engineering will construct stand-ins for our mental abilities—and possibly even enable us to fix our brains when they break.

## Brains and Their Construction

Brains, at one level, consist of ion channels, chemical pumps, specialized proteins. At another level, they contain several types of neurons connected via synaptic junctions. These are in turn composed into networks consisting of repeating modules of carefully arranged circuits. These networks are arrayed in interacting brain structures and systems, each with distinct internal wiring and each carrying out distinct functions. As in most complex systems, each level arises from those below it but is not readily reducible to its constituents. Our understanding of an organism depends on our understanding of its component organs, but also on the ongoing interactions among those parts, as is evident in differentiating a living organism from a dead one.

For instance, kidneys serve primarily to separate and excrete toxins from blood and to regulate chemical balances and blood pressure, so a kidney simulacrum would entail a nearly complete set of chemical and enzymatic reactions. A brain also monitors many critical regulatory mechanisms, and a complete understanding of it will include detailed chemical and biophysical characteristics.

But brains, alone among organs, produce thought, learning, recognition. No amount of engineering has yet equaled, let alone surpassed, brains' abilities at these tasks. Despite huge efforts and large budgets, we have no artificial systems that rival humans at recognizing faces, nor understanding natural languages, nor learning from experience.

There are, then, crucial principles that brains encode that have so far eluded the best efforts of scientists and engineers to decode. Much of computational neuroscience is aimed directly at attempting to decipher these principles.

Today we cannot yet fully simulate every aspect of a kidney, but we have passed a decisive threshold: we can build systems that replicate kidney principles so closely that they can supplant their function in patients who have suffered kidney loss or damage. Artificial kidneys do not use the same substrate as real kidneys; circuits and microfluidics take the place of cells and tissue, yet they carry out operations that are equivalent, and lifesaving, to the human bodies that use them. A primary long-term goal of computational neuroscience is to derive scientific principles of brain operation that will catalyze the comparable development of prosthetic brains and brain parts.

## Do We Know Enough About Brains to Build Them?

As with any complex system, in the absence of full computational understanding of the brain, we proceed by collecting constraints: experimentally observable data can rule out potential explanations. The more we can rule out, the closer we are to hypotheses that can account for the facts. Many constraining observations have usefully narrowed our understanding of how mental activity arises from brain circuitry; these can be organized into five key categories.

Brain component allometry: Remarkably tight relationships hold between a brain's overall size and the size of its constituent components. Just knowing the overall brain size of any mammal, we can with great precision predict the size of all component structures within the brain. Thus, with few exceptions, brains apparently do not and cannot choose which structures to differentially expand or reconfigure.[3–11] So, quite surprisingly, rather than a range of different circuits, or even selective resizing of brain components, human brains are instead largely built from the same components as other mammalian brains, in the same circuit layouts, with highly predictable relative sizes. Apparently a quantitative change (brain size) results in a qualitative one (uniquely human computational capabilities).[9, 12]

Telencephalic uniformity: Circuits throughout the forebrain (telencephalon) exhibit notably similar repeated designs,[6,13] with few exceptions,[14–19] including some slightly different cell types, circuit structures, and genes. Yet brain areas purported to underlie unique human abilities (e.g., language) barely differ from other structures; there are no extant hypotheses of how the modest observed genetic or anatomical differences could engender exceedingly different functions. Taken together, these findings intimate the existence of a few elemental core computational functions that are re-used for a broad range of apparently different sensory and cognitive operations.

Anatomical and physiological imprecision: Evidence suggests that neural components are surprisingly sloppy (probabilistic) in their operation, very sparsely connected, low-precision, and extraordinarily slow,[20–22] despite exhibiting careful timing under some experimental conditions.[23–27] Either brains are far more precise than we yet understand, or else they carry out families of algorithms whereby precise computations arise from imprecise components.[28–31] If so, this greatly constrains the types of operations that any brain circuits could be engaged in.

Task specification: Though artificial telephone operators field phone inquiries with impressive voice recognition, we know that they could do far better. The only reason we know this is that human operators substantially outperform them; there are no other formal specifications whatsoever that characterize the voice recognition task.[32,33] Engineers began by believing that they understood the task sufficiently to construct artificial operators. It has turned out that their specification of the task does not match the actual, still highly elusive set of steps that humans actually perform in recognizing speech. Without formal task specifications, the only way to equal human performance may be to come to understand the brain mechanisms that give rise to the behavior.

Parallel processing: Some recognition tasks take barely a few hundred milliseconds,[34,35] corresponding to no more than hundreds of serial neural steps (of milliseconds each), strongly indicating myriad neurons acting in parallel,[36] imposing a very strong constraint on the types of operations that individual neurons could be carrying out. Yet parallelism in computer science, even on a small scale, such as two or three simultaneous operations, has proven very elusive. Why, for instance, don't our dual-core or quad-core computers run two or four times faster than single-core systems? The (painfully direct) answer is that we simply do not yet know how to divide most software into parts that can effectively exploit the presence of these additional hardware elements. Even for readily parallelizable software, it is challenging to design hardware that yields scalable returns as processors are added.[37,38] It is increasingly possible that principles of brain architecture may help identify novel and powerful parallel machine designs.

## From Circuits to Algorithms to Prosthetics

There are several promising instances in which different laboratories (even laboratories that are competing with each other) have arrived at substantial points of agreement about what certain brain areas are likely doing. A notable success story arises from studies of the basal ganglia, which takes two kinds of inputs: sensory information from the neocortex, and "reward" and "punishment" information arising from external stimuli. We are close to computationally understanding this large chunk of the brain, which apparently carries out just one of our primary learning abilities: our slow "trial and error" learning (studied in computational neuroscience as "reinforcement learning"), underlying our ability to acquire such skills as riding a bike.[30, 39–65]

In addition, there is a growing consensus that circuits in the neocortex, by far the largest set of brain structures in humans, carry out another, quite different kind of learning: the ability to rapidly learn new facts and to organize newly acquired knowledge into vast hierarchical structures that encode complex relationships, such as categories and subcategories, episodes, and relations.[28, 66–74]

And these two systems are connected to each other, via far-reaching cortico-basal ganglia (aka cortico-striatal) loops .[49] The basal ganglia system carries out the computational operations of skill learning (reinforcement learning) while cortical circuits computationally construct vast hierarchies of facts and relations among facts. Interestingly, computational research on reinforcement learning has found that adding hierarchies to the process can greatly improve learning performance .[75,76] Our ancestors (reptiles and early mammals) were largely driven by the basal ganglia, whereas mammalian evolution has hugely expanded the relative size of the neocortex. By consistently increasing the size ratio of the neocortex to the basal ganglia, mammalian brain evolution may be solving a specific computational puzzle.[29, 75–79] Our understanding of human and animal learning abilities is being advanced by these computational studies, and we are developing novel methods for machine learning, enabling more powerful computer algorithms for analysis of complex data ranging from medical to commercial to financial applications.

Meanwhile, as study of these primary cortico-striatal brain structures remains very much still in progress, great advances have been made in deep, computational understanding of certain circumscribed brain systems, in particular those involved in early sensory transduction and perception. The results have been striking.

Analysis of cochlear mechanisms has led to the construction of prosthetics that serve today as cures for more than 100,000 people who have lost their hearing.[80] Retinal prosthetics are in advanced development.[81–85] In a recent study, patients with retinal implants recognized printed letters of size and distance comparable to reading a book in relatively low light. And experimental prosthetic arms can respond to brain-initiated control; people learn to control the arm simply by deciding to move it.[86, 87]

These sensory and motor findings have also led to formalizations of the general problem of acting in environments that are only partly observable and are dynamically changing, such as robotics or automated navigation; the result is a set of increasingly impressive robotic methods that see and navigate in complex surroundings.[88] In a series of trials run by the Department of Defense over the last several years, vehicles were, for the first time, able to navigate through real urban traffic, merging, passing, parking, and negotiating intersections, with no human control. Retinal algorithms operate equally well on other sensors such as radar; and prosthetic limb algorithms are wholly applicable to robots. Many of the algorithms that operate robots and automated vehicles are closely related to those that operate prosthetic limbs.

As we come to computationally understand how these peripheral sensorimotor systems work, the distinction between natural and artificial is being eroded. A breed of robots that share many of our own dexterity and perceptual abilities is likely to emerge directly from this research. As these increasingly biologically-based robots, or biots, come to replace human skilled labor, the economic and social consequences may be substantial.

## From Percept to Concept

The primary differences between human brains and those of other animals lie not in our sensory or motor mechanisms, which are largely shared across many species, but rather in cognitive abilities: association, representation, reasoning. Despite great advances in peripheral prosthetics, there is no commensurate understanding of advanced cognition.

The abilities of peripheral circuits (retina, cochlea, initial thalamic and cortical regions) are largely built in at birth via genetic programs and shaped in early childhood during developmentally critical periods. In contrast, the rest of the neocortex will use those built-in systems to acquire masses of specific information about the environment over a lifetime. Neocortical circuits are not born with knowledge of particular scenes, faces, or actions; these are acquired through sensorimotor experience: observing and interacting with objects and events in our surroundings. Cortical circuits are engaged almost entirely in fact learning: rapid, permanent acquisition and organization of everyday occurrences. The low-level biological mechanisms underpinning long-term fact learning (permanent, anatomical synaptic changes, rather than inherently ephemeral chemical changes) are becoming understood.[89] But the neocortex is not just a passive warehouse of billions of isolated facts; we can arbitrarily associate them, recall them, embellish them.[33] Association, recall, retrieval, organization—all that we can actually do with memory—depends on mechanisms that are as yet still unknown.

Early cortical areas, then, deal with recognizing objects (even in different lighting, settings, and clutter), but some laboratories are increasingly focusing on cortical circuits that are beyond the early sensory areas: the vast remainder of the neocortex that somehow encodes sequences, associations, and abstract relations.[33, 90–99]

Seeing a phone, we perceive not only its visual form but also its affordances (calling, texting, photographing, playing music), our memories of it (when we got it, where we have recently used it), and a wealth of potential associations (our ringtone, whom we might call, whether it is charged, etc.). The questions of how cross-modal information is learned and integrated, and in what form the knowledge is stored—how percepts become concepts—now constitute the primary frontier of work in computational neuroscience. In this borderland between perception and cognition, the peripheral language of the senses is transmuted to the internal lingua franca of the brain, freed from literal sensation and formulated into internal representations that can include a wealth of associations.

Even our simplest perceptions often rely on top-down processing: using stored memory representations to inform our ongoing perception and recognition. In some circumstances, we can recognize objects in just tens of milliseconds,[34,35] so rapidly that it is unlikely that any top-down pathways are yet engaged. Yet once we're beyond simple recognition, to the far richer range of inference, association, and even language, memories strongly influence our perceptions. Merely thinking of a car is sufficient to activate the same early visual areas that would have been triggered by actually seeing the car, including its shape, size, color, and other features.[100–102]

These early visual areas are just one instance of the spread of activation from a triggering memory.[103–105] Thinking of a car may also activate many other areas, as yet largely unmapped, that encode knowledge of how to open car doors, turn ignition keys, steer, accelerate, brake—or information about what particular car you own, where it is parked, and so on. Today we can experimentally test for visual shape information because we know a great deal about how to decode neural responses that occur in early visual areas,[106] but we have comparatively modest data for other associative knowledge.[107–109] Computational models of spreading activation[110,111] are now striving to make contact with specific neural mechanisms and brain pathways, to arrive at convergent hypotheses like those of peripheral sensory systems.

## Computing Individual Differences: From Neurotypes to Cognotypes

Though all of us have extraordinarily similar brains, even small differences can be striking. Whether particular characteristics are genetic, developmental, or learned is still often impossible to ascertain, but individual behavioral differences are highly likely to directly correspond to individual brain differences, whether genetic or acquired. Most work in computational neuroscience—from perception to cognition, from anatomy to computational models—has focused on one agent at a time, one brain at a time. A further frontier will be to confront differences among individuals.

Our bodies are built by genetic programs that became locked into particular patterns early on in mammalian evolution: four appendages; eyes above nose above mouth between ears; ten fingers and ten toes. We are not optimized to have just these features and no others; most of the variations that we might imagine—nose above eyes; five limbs; tentacles instead of hands—have never been tried by evolution, *not* because they're not improvements, but simply because they may not be consistent with the genetic prebuilt modules that have been bundled for hundreds of millions of years.[112,113]

Brain components are body components, so it is not surprising that evolutionary brain changes also are highly predictable, exhibiting selectional pressure only within the constraints of prescribed regularities: all mammals have almost exactly the same brain regions, in the same allometric size relationship, wired extraordinarily similarly.[5,6,9] It is hypothesized that the relatively modest brain differences that do occur are "canalized" into a relatively small set of categories. Brains create behaviors, and brain differences can create behavioral differences.

Because brain differences tend to follow certain patterns of architectural arrangements, or neurotypes, individual differences then tend to fall into groups, which correspondingly can be referred to as cognotypes. These can be described as a range of recognized characteristics of differential cognitive types,[114,115] such as types of psychopathy, personality attributes of introversion or extraversion, Asperger's characteristics, even different motor skill abilities. Any of these may entail a complex combination of inherited (genetic or polygenic) and acquired (developed or learned) characteristics. Symptoms of Asperger's, for instance, can include the seemingly arbitrary combination of high mathematical and engineering abilities with low social abilities, whereas there tend not to be behavioral types combining, say, high empathy with synesthesia, or low motor abilities with unusually high face-recognition abilities.

We won't achieve a full computational understanding of human brains until we understand why only certain variants tend to occur, and until we can model how it is that architectural brain differences can mechanistically generate cognitive differences. There are likely to be salient philosophical questions of will and intent, and ethical questions of capacity and culpability that will, it is hoped, be clarified as our understanding deepens.

## Extrapolations

The field of computational neuroscience is uniquely situated to decrypt the brain's mechanisms, to construe the perceptual and memorial abilities that still stymie our best engineering efforts. Once we crack the code, we finally will be able to construct systems that equal human performance at perceptual tasks. And having finally understood the underlying mechanisms, we may very well be able at long last to improve on them. There is no known formal reason why the capabilities of our brains may not eventually be equaled or exceeded.

There are economics to these advances, and policy implications abound. When auditory implants first became available, the scientific community widely doubted their efficacy. It took years of demonstration before they were accepted. They are expensive: today their cost can run to $100,000 per

patient. And there are risks: the surgical implantation procedure may lead to a higher incidence of meningitis.[116,117] Moreover, there are social complications: some in the deaf community find cochlear implants to be ethically misplaced, arguing that the deaf should not be thought of as disabled at all, but rather as a "minority cultural group."[118]

What of brain parts that are deeper than just the peripheral hearing system? Traumatic brain injury can cause debilitating deficits in memory and cognition; at present, such injuries are extremely difficult even to diagnose, let alone to treat. Implants to restore lost cognitive abilities for such accident victims would be revolutionary, and would be welcomed.

But if implants existed for accident-induced cognitive losses, could they also be used to augment uninjured cognitive function? There is suggestive evidence from drugs: some Alzheimer's medications may improve memory in people with mild cognitive impairment—but the FDA has not yet approved the use of any treatments for these lesser conditions.[119,120] How would regulators at the FDA react if it became possible to augment our brains—implants to help us think faster or to increase our memory capacity? The economic, social, and political concomitants of such technology would surely eclipse those arising from cochlear implants.

Each brain contains idiosyncrasies; our brains define who we are. The way we interact, the kinds of decisions we make, the connections we perceive—all arise from the still-obscure mechanisms of the vast span of thalamocortical circuits and cortico-striatal loops in our heads. These repeating components give us our mammalian abilities, our uniquely human faculties, and our individual characteristics. The computational understanding of individual and group differences will likely lead to a new science of different types of cognitive behavior, with implications ranging from law to education. The formerly familiar terrain of human nature may appear quite different in this light; perhaps, arriving there, we will truly know the place for the first time.

Our abilities are not inimitable; brain circuits are circuits, albeit nonstandard ones, and they will yield to analysis. As computational neuroscience comes to demystify them, we verge on an era of new frontiers in science and medicine, in which we can increasingly repair, enhance, and likely supplant the biological engines we think with.

**Richard Granger, Ph.D.**, is a professor at Dartmouth with faculty positions in the departments of psychological and brain sciences, computer science, and the Thayer School of Engineering. He directs Dartmouth's interdisciplinary Brain Engineering Laboratory, with research projects ranging from computation and robotics to neuroimaging and cognitive neuroscience. He has authored more than 100 scientific papers and holds numerous issued patents, is an elected fellow of the American Association for the Advancement of Science (AAAS), and serves on the boards of a number of technology corporations and government agencies. He is co-inventor of FDA-approved devices and drugs in clinical trials, and has been the principal architect of a series of advanced computational systems for military, commercial, and medical applications.

This work was supported in part by grants from the Office of Naval Research and from the Defense Advanced Research Projects Agency.